\documentclass[]{spie}  
\usepackage[]{graphicx}
\usepackage{hyperref}
\RequirePackage{abbrev_macros}
\def\bstfile{utphys}  

\title{Curious numerical coincidence to the Pioneer anomaly}

\author{Liviu Iv\u{a}nescu
\skiplinehalf {\it Earth and Atmospheric Sciences Department,
UQ\`AM, Montreal, QC, Canada}}

\authorinfo{
Send correspondence to Liviu Iv\u{a}nescu, ivanescu@sca.uqam.ca.\\
Address: D\'epartement des sciences de la Terre et de
l'Atmosph\`ere, Universit\'e du Qu\'ebec \`a Montr\'eal (UQ\`AM),
Case postale 8888, succursale Centre-Ville, Montreal, QC, H3C 3P8,
Canada.}

\begin{document}
  \maketitle

\begin{abstract}
One noticed a numerical coincidence between the Pioneer spacecrafts deceleration anomaly and $(\gamma-1)$, with $\gamma$ the Lorentz factor. The match is not only for distances larger than 20 AU, but even for the observed slop between 10 and 20 AU. Such numerical link may eventually lead to a scientific hypothesis for future theoretical investigations.
\end{abstract}


\section{INTRODUCTION}
\label{sec:intro}  
The Pioneer spacecrafts unmodeled deceleration (towards the Sun)
of $a_P = (8.74 \pm 1.33) \cdot 10^{-10}$ m/s$^2$, for heliocentric
distances greater than 20 AU, was reported in 1998
\cite{anderson-1998} and 2002\cite{anderson-2002}, based on an
initial dataset. This anomaly was confirmed\cite{turyshev-2009}
recently using newly recovered and carefully verified data. The
deceleration, observed for both Pioneer spacecrafts, was confirmed
by independent investigations \cite{markwardt-2002,
olsen-2007, levy-2009, toth-2009}. In this way, approximation
algorithms or errors in the navigation code have been ruled out as
possible causes of the anomaly. Alternatively, several physical
mechanisms came up and claimed being able to justify the target
value of $a_P$, by using standard or new physics
theories\cite{turyshev-2008}. For the time being, none seems to
convince the scientific community.

Another way to handle this issue is by reverse engineering, i.e.
finding an expression which gives a numerical coincidence to the
anomaly and only then try to build a model. One such example
is the fact that $\sqrt{G\cdot m_P/a_P}$ has the same order of
magnitude as the Compton wavelength of a proton, with $G$ the
gravitational constant and $m_P$ the proton
mass\cite{makela-2007}. Similarly, several authors explained why
$a_P \simeq c\cdot H_0$ could make sense\cite{tomilchik-2008},
with $H_0$ the Hubble constant and $c$ the speed of light in
vacuum.

Most of those models focus on a constant value, while it's not
certain that the anomaly is constant. Actually, the initial
dataset suggests\cite{anderson-2002} that the anomaly has
different values at distances less that 20 AU, while having a
small gradient towards the larger distances. In addition,
differences between the anomaly of Pioneer 10 and 11 could be
expected. The analysis of the newly recovered data may, hopefully,
clarify those aspects.

\section{Numerical Coincidence}

One proposes here a reverse engineering challenge starting with
the numerical coincidence that:
    \begin{equation}
    \label{eq:alpha}
a_P \simeq k\cdot(\gamma -1)\, ,
    \end{equation}
where $k=1$ m/s$^2$, $\gamma=1/\sqrt{1-\beta^2}$ is the Lorentz factor,
$\beta=v/c$, and $v$ is the radial spacecraft velocity with
respect to the Sun ($\sim 12$ Km/s at more than 20 AU). At a first
look, it could make sense that a residual value, as $a_P$, could be
explained by an excess factor, as $(\gamma-1)$, coming from the
Special Theory of Relativity. In addition, it's interesting to see how well it matches the observational values of $a_P$.

   \begin{figure}
   \begin{center}
   \begin{tabular}{c}
   \includegraphics[width=16.5cm]{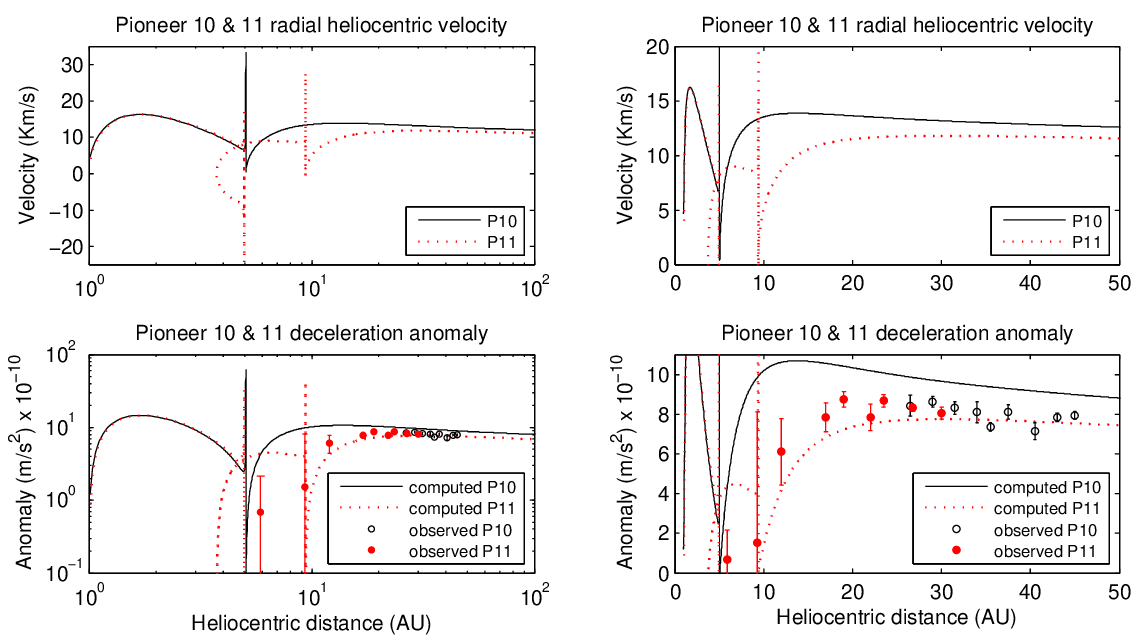}
   \end{tabular}
   \end{center}
   \caption[example]
   { \label{fig:anom}
Spacecrafts radial heliocentric velocity (upper graphs) and the
corresponding computed anomaly using $(\gamma -1)$ (lower graphs).
The left side graphs have logarithmic scales on the horizontal
axis, while on the right it's a zoom having linear scales.
Comparatively, the observed values of the
anomaly\cite{anderson-2002} with errorbars are presented. P10 and
P11 stand for Pioneer 10 and 11, respectively.}
  \end{figure}

As a first check, for $v=12$ Km/s, $(\gamma-1) = 8.0 \cdot 10^{-10}$, which
is very well in the range of the observed $a_P$ values. Secondly,
$(\gamma-1)$ varies as a fonction of $v$, which changes with the
heliocentric distance, and produces a very close match to the
observed $a_P$ (figure~\ref{fig:anom}). The trajectory data used
here comes from the JPL HORIZONS on-line solar system data and
ephemeris computation service\cite{jpl-2009}. The time stamps
corresponding to positions are ranging from few minutes to 7
hours, in order to provide smooth trajectories, especially during
Jupiter and Saturn flybys, at 5 and 9.4 AU, respectively.

Analyzing the figure~\ref{fig:anom}, one can observe that the
expression~(\ref{eq:alpha}) provides, for both spacecrafts, good
approximations to the observed $a_P$ values. Sometimes, $(\gamma
-1)$ is outside the errorbars, but the initial Pioneer dataset
also contained some bad values and therefore the errorbars may not
be very accurate. A lack of accuracy is suggested as well by the
fact that the observed values don't follow a very smooth trend, as
it should if they follow a certain model. One needs to emphasize
that the strong anomaly slop, between 10 and 20 AU, suggested by
the Pioneer 11 observations, is pointed out by $(\gamma -1)$
behavior too. Moreover, as the observed anomaly errorbars
represent the standard deviation over 10 days, this suggests
important variation of the anomaly at the Pioneer 11 Saturn flyby
(9.4 AU). The values given by $(\gamma -1)$ show such a behavior
too.

\section{Conclusions}

The reported Pioneer spacecrafts deceleration anomaly is computed from the observed trajectory. Here it was identified a curious numerical link between the observed deceleration and the relativistic term $(\gamma -1)$. This statement may eventually lead to a scientific hypothesis motivating a future investigation for a theoretical model explaining the Pioneer deceleration based on the $(\gamma -1)$ factor.

 \bibliographystyle{\bstfile}
 \bibliography{pioneer}

\end{document}